\documentclass[12pt,epsf]{article}
\usepackage{graphicx}
\usepackage{epsfig}
\usepackage{amsmath}
\usepackage{subfigure}
\usepackage{slashbox}
\begin{document}

\def\a{\alpha}
\def\b{\beta}
\def\c{\varepsilon}
\def\d{\delta}
\def\e{\epsilon}
\def\f{\phi}
\def\g{\gamma}
\def\h{\theta}
\def\k{\kappa}
\def\l{\lambda}
\def\m{\mu}
\def\n{\nu}
\def\p{\psi}
\def\q{\partial}
\def\r{\rho}
\def\s{\sigma}
\def\t{\tau}
\def\u{\upsilon}
\def\v{\varphi}
\def\w{\omega}
\def\x{\xi}
\def\y{\eta}
\def\z{\zeta}
\def\D{\Delta}
\def\G{\Gamma}
\def\H{\Theta}
\def\L{\Lambda}
\def\F{\Phi}
\def\P{\Psi}
\def\S{\Sigma}

\def\o{\over}
\def\beq{\begin{eqnarray}}
\def\eeq{\end{eqnarray}}
\newcommand{\lsim}{\raisebox{0.6mm}{$\, <$} \hspace{-4.4mm}\raisebox{-1.5mm}{\em$\sim\,$}}
\newcommand{\gsim}{\raisebox{0.6mm}{$\, >$} \hspace{-3.0mm}\raisebox{-1.5mm}{\em $\sim\,$}}
\newcommand{\vev}[1]{ \left\langle {#1} \right\rangle }
\newcommand{\bra}[1]{ \langle {#1} | }
\newcommand{\ket}[1]{ | {#1} \rangle }
\newcommand{\EV}{ {\rm eV} }
\newcommand{\MeV}{ {\rm MeV} }
\newcommand{\GeV}{ {\rm GeV} }
\newcommand{\TeV}{ {\rm TeV} }
\def\diag{\mathop{\rm diag}\nolimits}
\def\Spin{\mathop{\rm Spin}}
\def\SO{\mathop{\rm SO}}
\def\O{\mathop{\rm O}}
\def\SU{\mathop{\rm SU}}
\def\U{\mathop{\rm U}}
\def\Sp{\mathop{\rm Sp}}
\def\SL{\mathop{\rm SL}}
\def\tr{\mathop{\rm tr}}

\def\IJMP{Int.~J.~Mod.~Phys. }
\def\MPL{Mod.~Phys.~Lett. }
\def\NP{Nucl.~Phys. }
\def\PL{Phys.~Lett. }
\def\PR{Phys.~Rev. }
\def\PRL{Phys.~Rev.~Lett. }
\def\PTP{Prog.~Theor.~Phys. }
\def\ZP{Z.~Phys. }

\def\Z{\mathcal{Z}}
\def\W{\Omega}

\def\stau1{\tilde{\tau}_{1}}
\def\bino{\tilde{B}}
\def\n1{\tilde{\chi}^{0}_{1}}


\baselineskip 0.7cm

\begin{titlepage}

\begin{flushright}
UT-11-09\\
IPMU-11-0059\\
KEK-TH-1452\\
\end{flushright}

\vskip 1.35cm
\begin{center}
{\large \bf
A Possible Interpretation of CDF Dijet Mass Anomaly and \\
its Realization in Supersymmetry
}
\vskip 1.2cm
Ryosuke Sato$^{1,2}$, Satoshi Shirai$^3$ and Kazuya Yonekura$^{1,2}$
\vskip 0.4cm

{\it
$^1$Department of Physics, University of Tokyo, 
Tokyo 113-0033,
Japan\\

$^2$IPMU,
University of Tokyo, 
Chiba 277-8586, 
Japan\\

$^3$ Institute of Particle and Nuclear Studies,\\
High Energy Accelerator Research Organization (KEK)\\
Oho 1-1, Tsukuba, Ibaraki 305-0801, Japan\\
}

\vskip 1.5cm

\abstract{Recently, the CDF collaboration reports an anomaly in dijet mass distribution in association with a lepton and missing energy.
We discuss a possibility that the origin of the lepton and missing energy comes not from a $W$ boson but a new boson particle, which
is also responsible for  the dijet mass peak.
We show that such a situation can be realized in the framework of the minimal supersymmetric standard model and the dijet anomaly can be explained. 
}
\end{center}
\end{titlepage}

\setcounter{page}{2}

\section{Introduction}

Recently the CDF collaboration has reported the peak of the dijet mass round 140 GeV \cite{Aaltonen:2011mk}.
The search is performed in dijet events
with one lepton + missing energy,
motivated with the production of a new particle in association with a $W$ boson,
\beq
p{\bar p}\to \cdots \to W^{\pm}(\to \ell^{\pm}\nu) + \phi(\to{\rm 2~jets}).
\eeq
Here, $\ell^\pm$ is an electron or muon.
Such a mass peak is difficult to be explained in the framework of the Standard Model (SM).
The SM Higgs particle has too small cross section to explain the mass peak.
Therefore we expect a new source beyond the SM (BSM).

There are many researches done on this topic \cite{Bai:2010dj,Buckley:2011vc,Yu:2011cw,Eichten:2011sh,Kilic:2011sr,Wang:2011uq,Cheung:2011zt}.
Usually, the origin of the lepton and missing energy is assumed to be a leptonic decay of a $W$ boson. 
However, we can also consider the case that the source of the lepton and the missing energy is not the $W$ boson decay.\footnote{
This possibility is also pointed out in Ref. \cite{Yu:2011cw}}
Here, we consider an alternative case that the source of the lepton + missing energy comes from
a charged and color singlet particle, which is also responsible for the dijet mass peak.
Let us consider the production of the new particles
\beq
p{\bar p} \to \cdots \to \phi^{\pm}(\to \ell^{\pm}\nu) +\phi^{\mp}(\to qq'), \label{eq:process}
\eeq
where $\phi$ is the new boson particle.
The leptonic decay of $\phi$ provides the lepton and missing energy.

If the cross section of the process is order of pb, it is possible to explain the dijet anomaly.
Considering the direct production, if the interaction strength of the $\phi$ is of the same order as the electroweak interaction,
it is difficult to achieve the preferred cross section in the case that the $\phi$ is a scalar particle 
and it would be possible in the case of the vector $\phi$.
Therefore the case of massive vector $\phi$ such as a $W'$ boson is an interesting possibility to explain the dijet anomaly.

However, it is not necessary that the $\phi$'s are directly produced, and
in that case the boson $\phi$ can be a scalar field.
We assume the scalar particle is produced from the cascade decay of a slightly heavier particle $f$, which has a large production cross section:
\beq
p{\bar p} \to f(\to X \phi)+f'(\to X' \phi). \label{eq:process2}
\eeq
If the decay products $X, X'$ from $f$ are high energy objects such as jets or leptons,
the signal acceptance is reduced and 
this scenario might be inconsistent with existing BSM searches.
Therefore $X$ should be  low energy objects or neutrinos.
Such a situation can be realized in many BSMs, including supersymmetric (SUSY) models as we will discuss.

\section{Test of two scenarios}
In this Letter, we consider the case that the lepton and missing energy comes from not a $W$ boson but a new boson particle $\phi$.
Here we propose a simple test to discriminate between the two scenarios.

\begin{figure}
\begin{center}
\includegraphics[width=10cm]{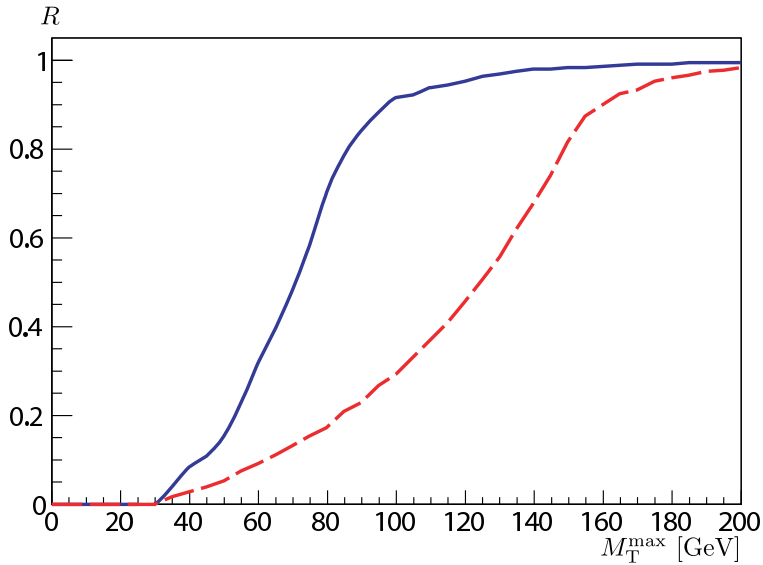}
\end{center}
\caption{The reduction factor $R$ (defined in the text) as a function of $M_{\rm T}^{\rm max}$.
The blue and solid line represents the case of usual $W$ boson mass and
the red and dashed line represents the case of unusual $W$ boson mass 150 GeV.
}
\label{R}
\end{figure}

In the paper of the CDF, the upper-bound on the transverse mass $M_{\rm T}(\ell \nu)$ is not imposed.
However, if the origin of the lepton and missing energy is not a $W$ boson, the transverse mass distribution is changed.
The $M_{\rm T}(\ell \nu)$ distribution from the $\phi$ decay has a Jacobean peak at not $W$ mass but $\phi$ mass,
if $\ell$ is not a tauon.~\footnote{
When $\ell$ is a tauon, the Jacobean peak are smeared.}
Here we assume that the particles from the decay of $f,~f'$ in Eq.~(\ref{eq:process2}) other than $\phi$ have negligible energy due to degeneration of the masses of $f,~f'$ and $\phi$.
Therefore, by imposing an additional cut: upper bound on $M_{\rm T}(\ell \nu)$,
the two scenarios would be easily discriminated.
In order to confirm this, we consider the process of a Higgs production in association with a $W$ boson:
$p{\bar p} \to Wh$.
We set the mass of $h$ to be 150 GeV.
We consider two cases: the usual $W$ boson mass and the mass changed to 150 GeV.
The events are generated with the program Pythia 6.4 \cite{Sjostrand:2006za}.
The detector simulation is done with the program AcerDet \cite{RichterWas:2002ch} with modification to adjust for the Tevatron detector.
Then we impose the similar event selection as in the CDF  study  \cite{Aaltonen:2011mk}.
Additionally we impose the upper-bound   on $M_{\rm T}(\ell \nu)$.
We define the reduction factor $R$ as
\beq
R(M_{\rm T}^{\rm max}) \equiv \frac{{\rm \#~ of~ events~ with~additional~cut~} M_{\rm T}<M_{\rm T}^{\rm max}}{ {\rm \#~ of~ events}}.
\eeq
It is expected that the $R$ is small in the case of $\phi$.
In Fig. \ref{R}, we show the plot of $R$ as a function of $M_{\rm T}^{\rm max}$.

We can see that the value of $R$ in the case of the heavy $W$ boson is much smaller compared to the normal $W$ boson case around $M_{\rm T}^{\rm max}=100 $ GeV.
Therefore, by imposing the additional cut on $M_{\rm T}$, the height of dijet mass peak is drastically reduced.
Equivalently, a tight cut of a lower bound on $M_{\rm T}$, e.g. $M_{\rm T}>100~{\rm GeV}$ will reduce the background if the source of the lepton + missing energy
is the boson $\phi$.

\section{Realization in SUSY Model}
In this section, we show that the dijet anomaly can be explained in the framework of the minimal SUSY models and
discuss the constraints from existing results from collider experiments and
future prospects at the LHC.
\subsection{Model Setup}
We consider the SUSY model in which the wino is the lightest neutralino and chargino and
the slepton is the lightest SUSY particle (LSP), which can be achieved e.g. in anomaly mediation models~\cite{Randall:1998uk,Giudice:1998xp}.
In addition, we consider a slight violation of the R-parity (see Ref.~\cite{Barbier:2004ez} for a review).
In general, the R-parity violating superpotential is written as
\begin{eqnarray}
\label{Rviolation}
W_{R\hspace{-0.40em}/}=\frac{1}{2}\lambda_{ijk}L_iL_j\bar{E}_k
+\lambda^{\prime}_{ijk}L_{i}Q_{j}\bar{D}_k
+\frac{1}{2}\lambda^{\prime\prime}_{ijk}\bar{U}_{i}\bar{D}_{j}\bar{D}_k +
\varepsilon_i L_i H_u.
\end{eqnarray}
For simplicity, we set $\varepsilon=0$.
We assume that there are no baryon number violating terms, that is, $\lambda''=0$.
We also assume that the lepton flavor violation in the first and second generation is negligible, and 
only consider the violation of the lepton number in the third generation, i.e., $\lambda_{3ii}, \lambda'_{3ij} \neq 0$.
In this setup, we can avoid~\cite{Davidson} the wash-out of the baryon asymmetry in early universe~\cite{RpVcosmobound}
if the lepton flavor is not strongly violated in the soft masses~\cite{Endo:2009cv}.

In this case the LSP slepton is no longer stable.
The slepton decays into lepton+neutrino or two quarks.
By adjusting the parameters $\lambda$ and $\lambda'$, it is possible to realize
${\rm Br}(\tilde{\ell} \to \ell \nu) = {\rm Br}(\tilde{\ell} \to qq')=0.5$ and in this case, the signal acceptance gets maximum.

\begin{figure}
\begin{center}
\includegraphics[width=10cm]{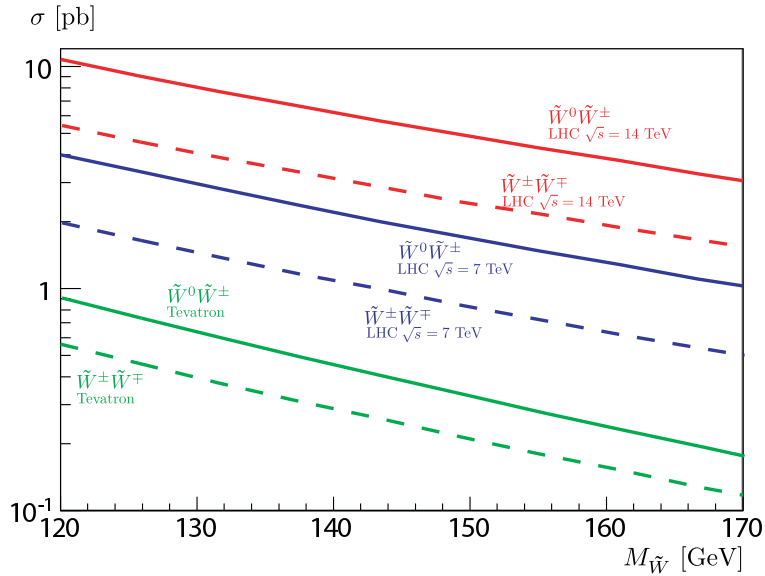}
\end{center}
\caption{The NLO production cross section of the winos at the Tevatron and the LHC.
We have used the program Prospino2 \cite{Beenakker:1996ed}.
}
\label{cross}
\end{figure}

We consider the process
\beq
p{\bar p} \to \tilde{W}\tilde{W} \to \tilde{\ell}^{\pm}(\to \ell^{\pm}\nu) +\tilde{\ell}^{\mp}(\to qq'), \label{eq:process3}
\eeq
In the case that $m_{\tilde{W}}=140-150$ GeV, the cross section is about 0.5 pb as shown in Fig. \ref{cross}
 and it is possible to explain the CDF dijet anomaly. The slepton mass is also assumed to be around $140-150$ GeV to explain
 the dijet mass peak. Then the $\tilde{W}$ and $\tilde{\ell}$ are almost degenerate and hence we can neglect the SM particles (neutrino or lepton) coming from 
 the decay of $\tilde{W}$.  

For concreteness, we perform the analysis in the case of the minimal anomaly mediation.
We set the model parameters as
\beq
m_0=292~ {\rm GeV}, ~m_{3/2}=51~ {\rm TeV},~{\rm sign}(\mu)=+1, ~ \tan\beta=10.
\eeq
The mass spectrum and decay table are generated by the ISAJET 7.80 \cite{Paige:2003mg}.
The physical mass parameters are
$m_{\tilde{\chi}^0_1}=146$ GeV, $m_{\tilde{\chi}^\pm_1}=147$ GeV and $m_{\tilde{\tau}_1}=144$ GeV.
The branching fraction of the R-parity violating LSP decay is calculated with the program Pythia 6.4.
We set $\lambda_{311}=\lambda_{322}=1.2\times 10^{-4}$ and $\lambda'_{312}=10^{-4}$.
Because of these R-parity violated couplings, the LSP $\tilde{\tau}_1$ decays with the branching ratios 
${\rm Br}(\tilde{\tau}_1 \to e\nu)={\rm Br}(\tilde{\tau}_1 \to \mu\nu)=0.25$ and 
${\rm Br}(\tilde{\tau}_1 \to us)=0.5$.
The decays of the winos are $\tilde{\chi}^0_1 \to \tau^{\pm}\tilde{\tau}^{\mp}_1$ and 
$\tilde{\chi}^\pm_1 \to \nu \tilde{\tau}^{\pm}_1$.
The energy of the tauon from the wino decay is very small because of the smallness of the mass difference and 
the tauon is irrelevant for the following analysis.
At the Tevatron, the production of SUSY particles are dominated by lightest chargino and neutralino production.
The NLO cross section is calculated by the program Prospino and is 0.63 pb.

We impose the similar event selection as in Ref. \cite{Aaltonen:2011mk}.
There are some detector effects which cannot be treated by the fast detector simulation such as particle identification efficiency,
which reduce the number of the signals.
To estimate the reduction factor, we generated diboson process at NLO level and compared to the result of simulation in Ref. \cite{Aaltonen:2011mk}. 
We multiply the SUSY cross section by the overall reduction factor.

\begin{figure}
\begin{center}
\includegraphics[width=10cm]{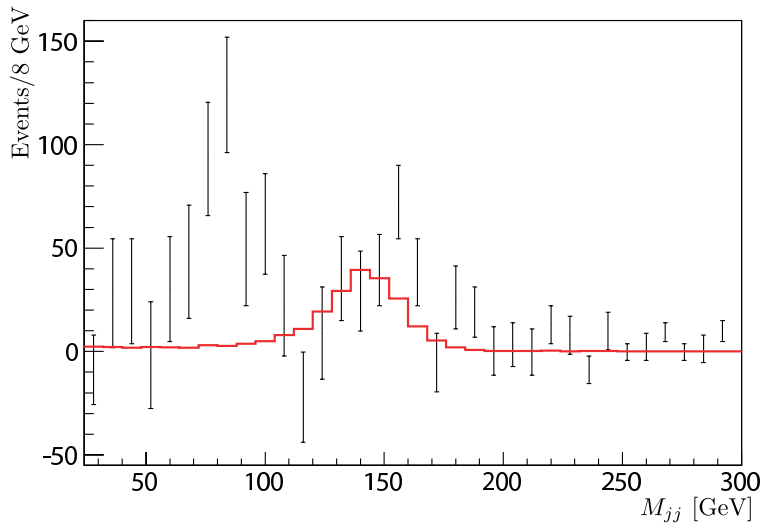}
\end{center}
\caption{The dijet mass distribution in the present model.
Black data points show the observed resonant contribution to $M_{jj}$ in Ref. \cite{Aaltonen:2011mk} including $WW$ and $WZ$ production.}
\label{dijet}
\end{figure}

In Fig. \ref{dijet}, we show the dijet mass distribution in the present model.
One can see that this model can explain the dijet mass peak.

Finally let us comment on the choice of R-parity violating parameters.
The present R-parity violating couplings are consistent with phenomenological constraints~\cite{Barbier:2004ez,RpV-bounds}.
$\l_{ijk}$ are tuned for lepton flavour, because the excess is observed in both electron and muon channel. 
We need to set the parameters such that $\l_{311} \simeq \l_{322}$.
We also switched off couplings which leads to ${\tilde\tau}\to\tau\nu$ because hadronic decay of $\tau$ reduces signal acceptance.
For example, we need to assume that $\l_{313}$ is negligible.
As for $\l'_{ijk}$, there are freedom for the choice of flavor structures, as long as the hadronic branch and leptonic branch of $\tilde{\tau}$ is of the same order
to maximize the signal acceptance.
We also mention about the Grand Unified Theory (GUT).
The R-parity violating Yukawa couplings we have introduced is not $SU(5)$ GUT invariant.
However, the Yukawa coupling unification may not necessarily maintained for small Yukawa couplings (e.g., by using the GUT breaking field VEV),
and hence it is not so obvious whether the present choice of the parameters immediately contradicts with the GUT or not.

\subsection{Constraints and Prospects}
In addition to the present one lepton+missing+dijet signature, 
there are different patterns of the final states, depending on the decays of the slepton LSP.
Possible final states are:
\begin{description}
\item[Case 1.] One lepton+missing energy+dijet
\item[Case 2.] Same sign (SS) 2 leptons + missing energy 
\item[Case 3.] Opposite sign (OS) 2 leptons + missing energy
\item[Case 4.] Dijet+dijet
\end{description}
Each case occurs at  frequency comparable to each other.
In the presence of production of the heavier SUSY particles such as a gluino and a squark,
there are additional signatures such as high $P_{\rm T}$ jets.
The Case 2. and 3. are very similar to the signatures of low-energy gauge mediation in which
the wino production is dominated.
Therefore, studies on the constraint \cite{Ruderman:2010kj} and search \cite{Nakamura:2010faa}
 on the low-energy gauge mediation are almost straightforwardly applicable.
The search based on the Case 4. would be challenging because of large amount of QCD background. 
In this section, we discuss the constraint on the present model.
\subsubsection*{Constraint}
The strongest constraint comes from SS lepton + missing energy signature, which is almost free from the SM background.
The CDF collaboration provides data of the same sign 2 leptons and missing energy event \cite{Abulencia:2007rd}.
They have observed 13 events after some cuts and the SM prediction is $7.8\pm 1.1$.
At the model point presented above, the predicted number of events is about 5.
Therefore the current model point is consistent with the CDF SS2l search.
Recently, the ATLAS collaboration has reported the analysis on two leptons + missing energy events \cite{Collaboration:2011xm}.
They have observed 0 events with the same sign 2 leptons + missing energy events, which is consistent with the SM prediction 0.28 $\pm$ 0.14.
On the other hand, the present model predicts about 1 event.
Thus, we conclude that this model is consistent with the ATLAS search.
The search for opposite sign 2 leptons event gives weaker constraint on the present model.
\subsubsection*{Prospects}
As discussed previously, the SS2l search is the most promising search for the present model.
In the year 2011, the integrated luminosity of the LHC  will reach ${\cal O}(1) ~{\rm fb}^{-1}$.
In this case, it is easy to discover the present model.

In the above discussion, we consider the wino production (see Fig. \ref{cross}).
In the framework of the anomaly mediation,
the gluino and squark masses are predicted to be around 1100 GeV.
For integrated luminosity of a few ${\rm fb}^{-1}$,  signals from such colored particles can also be expected.
Such signals are accompanied with high $p_{\rm T}$ jets besides multi-lepton and missing energy.
Observation of such signals is a crucial test for the present model.

\section{Conclusion and Discussion}
In this Letter, we tackle the CDF dijet mass anomaly in  lepton + missing energy events.
We propose a scenario that the origin of a lepton and missing energy is not a $W$ boson but a new
BSM particle, which also makes the dijet mass peak.
We discuss a simple test for this scenario using the lepton transverse mass.
In addition, we show that such a scenario can be realized in SUSY models.
This model is consistent with the current experimental data and
can be tested with the very near future LHC run.

\section*{Acknowledgements}
We would like to thank T. T. Yanagida for useful comments, and
M. Endo and T. Moroi for useful discussions.
This work is supported in part by JSPS
Research Fellowships for Young Scientists and by
World Premier International Research Center Initiative, MEXT, Japan.


\end{document}